# Modulation of diffusion rate of therapeutic peptide drugs using graphene oxide membranes.

T. M. Puvirajesinghe[1,2,3*], Z.L. Zhi[4], R. V. Craster[5], and S. Guenneau[3,6,7].

[1]CRCM, Inserm, U1068, Marseille, F-13009, France; [2]Institut Paoli-Calmettes, Marseille, F-13009, France; [3]Aix-Marseille Université, F-13284, Marseille, France. [4]Diabetes Research Group, King's College London Faculty of Life Sciences and Medicine, Guy's Hospital, London, United Kingdom; [5]Department of Mathematics, Imperial College London, United Kingdom; [6]Centrale Marseille, 38 Rue Frédéric Joliot Curie, F-13013 Marseille, France; [7]Institut Fresnel, CNRS, UMR 7249, Avenue Escadrille Normandie Niemen, F-13013, Marseille, France.

*Corresponding author.

KEYWORDS: graphene oxide; peptide drugs; mass diffusion; finite elements.

**We investigate diffusion of a peptide drug through Graphene Oxide (GO) membranes that are modeled as a porous layered laminate constructed from aligned flakes of GO. Our experiments using a peptide drug show a tunable non-linear dependence of the peptide concentration upon time. This is confirmed using numerical simulations with a diffusion equation accounting for the photothermal degradation of fluorophores and an effective percolation model. This modeling yields an interpretation of the control and delay of drug diffusion through GO membranes. The ability to modulate the density of hydrogel-like GO**

**membranes to control drug release rates could be a step forwards in tailoring drug release properties of the hydrogels for therapeutic applications.**

Graphene, a single-atom thick free-standing layer of graphite, has been studied in various physical and chemical contexts. It is a single-layer honeycomb lattice of carbon atoms consisting of two interlocking triangular lattices, which has been extensively studied in various physical and chemical contexts[1,2]. The present work explores the diffusion-type features in graphene through an effective medium approach. However, the electronic properties of graphene underpin the physics at the nanoscopic scale and we find it worthwhile recalling related results to understand the basis of our modelling approach of diffusion of a peptide drug through a GO membrane. Well-known experiments on graphene include the demonstration that its electrical conductivity as a function of charge density increases symmetrically on either side of a minimum value at the neutrality point[3]. It is now believed that minimum conductivity might be an artifact of extrinsic electrons and hole puddles for which similar theoretical conductivity estimates exist on the basis of percolation and tunneling between adjacent puddles[4]. Other properties of graphene underpinned by diffusion and percolation models include filtration applications of water molecules[5] at the nanometric scale within a GO flake. Inspired by all these works we investigate, theoretically and experimentally, a path towards control and delay of drug release with hydrogel-like GO membranes.

We have previously adapted the concept of transformation thermodynamics[6], whereby the flux of temperature is controlled via anisotropic heterogeneous diffusivity, for the diffusion and transport of mass concentration. The $n$-dimensional, time-dependent, anisotropic heterogeneous Fick's equation was considered[6], which is a parabolic partial differential equation applicable to heat diffusion, for instance, in fluids. Finite-element computations were initially used to model liposome particles surrounded by a spherical multi-layered cloak consisting of layers of fluid with an isotropic homogeneous diffusivity, deduced from an effective medium approach[6]. This theoretical model has been experimentally validated for chemical engineering, whereby a steel structure in a concrete foundation can be protected from seawater corrosion by surrounding the structure with substances according to their anisotropic heterogeneous diffusivity[7]. Such an effect of invisibility in chemical engineering was inspired by earlier works in optics and acoustics[8-12]. Other applications include invisibility cloaks controlling light diffusion in water-based media[13]. We now investigate how the concept of anisotropic homogenous diffusion can be applied to tune drug delivery. Peptide and protein drugs have serum half-lives of minutes to hours, however conjugation to polymers in hydrogels results in the retardation of kidney filtration and a corresponding increase in plasma half-life of the therapeutic drug[14]. Clinical advantages include fewer injections for patients and reduced side effects for the health of patients. Based on the homogenization model applied in Guenneau and Puvirajesinghe, 2013[6], GO was selected because of its low permeability[15]. GO has already been described to be a molecular sieve as the GO laminates only allow the permeation of ions of a certain hydrated radius[5,16]. GO has been used in combination with other composite materials to control drug release within hydrogels[17-19]. This study demonstrates similarities with GO membranes and clay and clay-based materials. Indeed this explains the combination of the two materials for the fabrication of clay-graphene composites[20-22]. To model GO hydrogel-like membranes, we employ a porous layered laminate to describe the percolation of an aqueous medium in these structures[23,24]. In combination with numerical stimulations, the concentration of GO in hydrogels can be used as a parameter to vary the rate of diffusion of a therapeutic peptide drug in order to achieve tunable drug release. Using these principles, one can fine-tune the drug release properties of GO hydrogels by calculating the overall GO composition.

## Results

**GO membranes and measurement of the diffusion rate of a fluorescence anticancer lytic peptide**

GO has already been employed in hydrogels and provides advantages such as water solubility, high specific surface area and good biocompatibility. The anti-cancer agent used in this study is a previously characterized cationic lytic peptide, whose mechanism of action was based on disintegrating the cell membrane, leading to cell death. Incubation of MDA-MB-231 (human breast basal epithelial cancer cells) with different concentrations of the peptide has been reported to show a dose-dependent reduction in cell proliferation[25]. The peptide drug was chemically synthesized with the addition of a fluorescein isothiocyanate (FITC) fluorophore at the N-terminal of the peptide. This enabled the monitoring of the presence of the peptide using green fluorescence. The concentration of the peptide was proportional to the fluorescence intensity and calibration curves can be derived to measure peptide drug concentration (**Supplementary Fig. S1**).

In order to study the diffusion rate of the fluorescently labelled drug, we used cell culture membrane inserts made of polyethylene membrane with a porosity of 0.4 microns. The GO was deposited onto the membrane itself (**Fig. 1a**) and dried by exposure to high heat. A fluorescent plate reader was used to monitor the rate of drug diffusion by measuring the accumulated fluorescence signal from the lower chamber (**Fig. 1b, c** and **d**). Automatic readings were taken from the bottom of the plate at regular intervals of 3 min for 12 h 45 min (**Fig. 2**). In the absence of the GO membrane, the time taken for half the maximum amount of peptide to pass through the membrane was 18 min. Increasing the GO membrane density retards, and eventually halts, the rate of transport (see Supplementary Fig. S2). In comparison to no GO membrane the density increases to 0.06, 0.09 and 0.17 ng/mm$^3$ retard the peptide drug by 4, 6, 7 fold and completely, respectively.

Curves depicting the diffusion of the drug (**Fig. 2**) showed that the magnitude of GO density affected the rate of drug transport. Depositing increasing densities of GO lead to a gradual decrease in the time taken for the drug to reach half of its maximal intensity ($t_{1/2}$). Therefore the density of GO determines the rate of retardation of drug transport.

In order to interpret this phenomenon, the membrane compactness of GO was characterized using transmission electron microscopy (TEM). Based on the fact that GO consists of a high proportion of carbon atoms, for which the low atomic number reduces scattering of the electron beam, a good contrast and a sharp image was acquired in TEM experiments without the need of the addition of a contrast agent. Low GO densities show isolation of GO structures (**Fig. 3a**). When the density of the GO membrane increased, TEM images showed that the heterogeneous GO have the tendency to aggregate, causing overlap of GO structures (**Fig. 3b**). However upon reaching a certain concentration of GO, complete overlap was evident, which leads to superimposition of particulates, as evidenced by an overall dark grey image, shown in (**Fig. 3c**). Therefore in conclusion, there is strong evidence to validate our proposal of the application of GO for the tunable delivery of therapeutic drugs. Next, we compare the acquired experimental results with effective mathematical models, which showed a close match between the numerical data and the experimental data (**Supplementary Fig. S3**).

For the numerical and analytical simulations, we recall that the assembly of aligned GO flakes resembles a porous layered laminate. It has been previously shown that a homogenization approach of Fick's equation can be used to control processes of mass diffusion for biological and engineering applications[6]. We use the same equation to retrieve in (**Fig. 4)** the main features of the experiments reported in (**Fig. 2)**. However, one notes that there is a decrease in the concentration after a certain time in (**Fig. 2)**, whereas (**Fig. 4c)** only displays a plateau. We explain how this can be corrected by considering certain activation terms in the diffusion equation so as to account for photothermal degradation of fluorophores, described in the later section on methods.

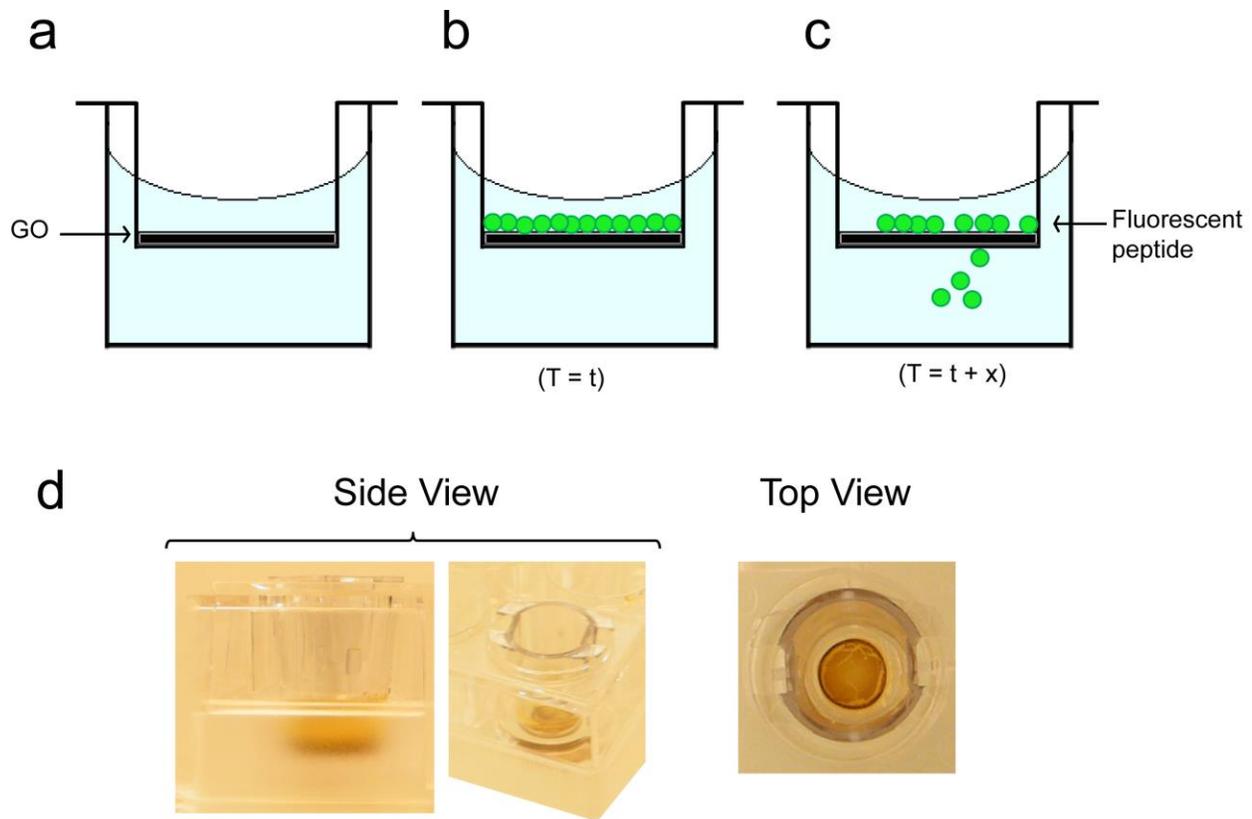

**Fig. 1: Experimental setup for the measurement of the rate of diffusion of a fluorescence anticancer lytic peptide.** **(a)** Schematic representation of graphene oxide (GO) deposited onto the polyethylene terephthalate (PET) membrane. **(b)** A fluorescent peptide with anti-cancer lytic activity is added to the Transwell cell culture insert at the beginning of the experiment (T=t). **(c)** At increasing time points during the experiment (T=t+x), the concentration of fluorescent peptide in the lower culture well chamber is measured. **(d)** Side view and top view photographs of the experiment depicted in (a,b,c).

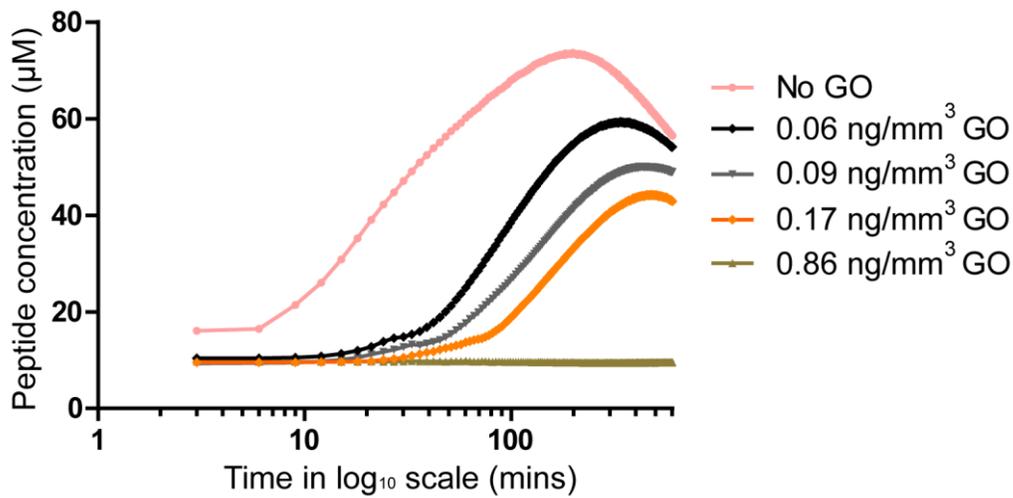

**Fig. 2: Varying the density of the GO membranes controls the rate of diffusion of an anticancer lytic peptide.** GO dispersion in water is prepared at different densities and dried onto translucent cell culture inserts made from PET membrane, with a porosity of 0.4 microns. The concentration of the fluorescent peptide in the lower culture well chamber is measured and plotted against time. Each reading is representative of an average measurement of 24 readings taken from the circumference of each well of the culture dish. Readings are taken every 3 min during a period of 12 h 45 min.

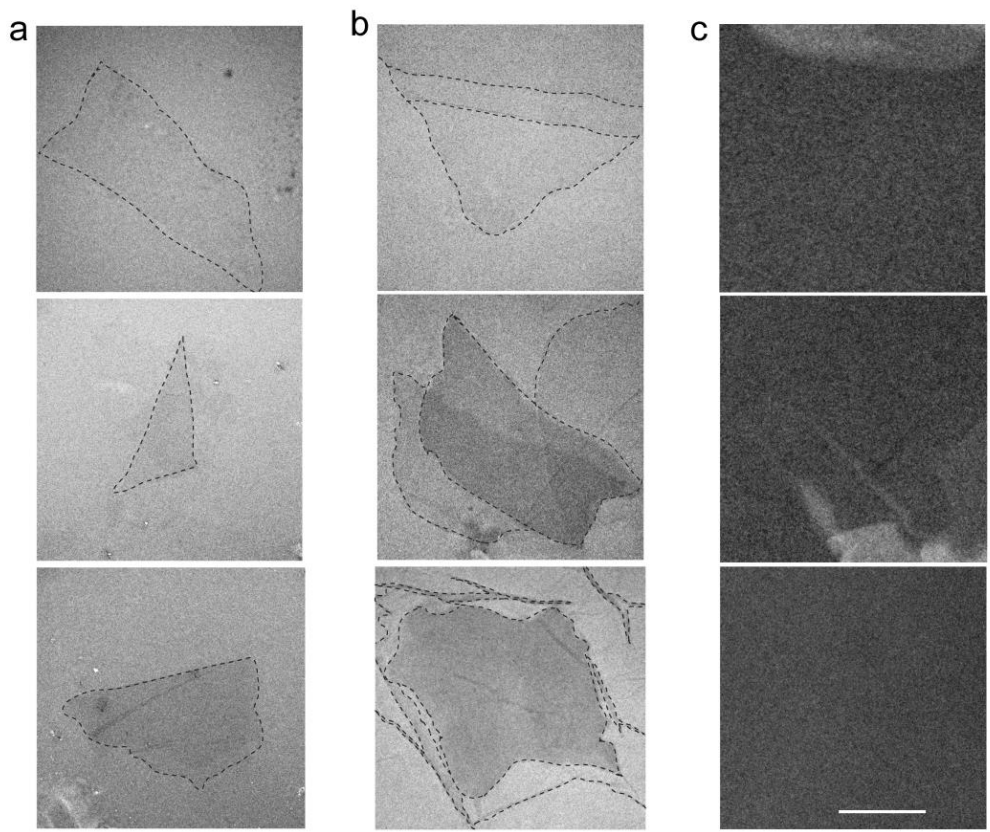

**Fig. 3: TEM images showing GO of varying densities**. Images are acquired using a Morgagni FEI 80KV Camera digital View III Olympus camera. For clarity the prominent sheets are outlined with dashed lines. (**a**) An image of one sheet, note that the contrast is very similar to that of the background surface. (**b**) Several sheets, each sheet is again outlined and the contrast is increased when additional sheets are superimposed. (**c**) Multiple sheets showing a sharp increase in the contrast when many GO flakes are stacked together, thereby substantially increasing the gray scale of the images. Scale bar (500 nm).

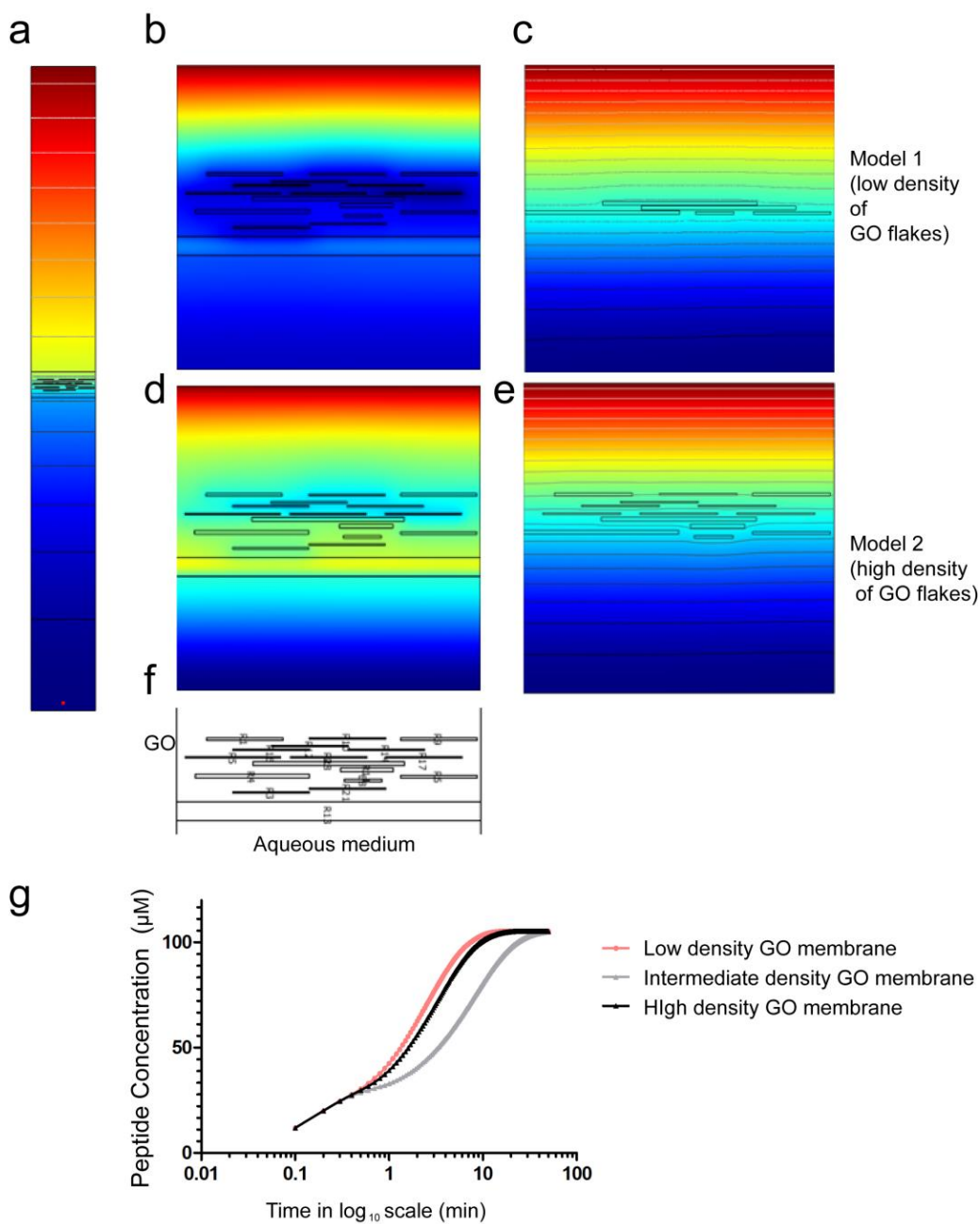

**Fig. 4: Finite element simulations for effective diffusion rate.** (**a**) A long water-filled rectangular domain with no flux boundary conditions on the left, right and bottom sides; an imposed concentration on the top and GO flakes overlying a porous (PET) membrane. (**b,d**) Concentration of the peptide at time steps t=1min and t= 200 min. (**c,e**) Concentration of the peptide for low and high density of GO flakes. (**f**) Geometry of PET overlaid with a high density of GO. (**g**) Variation of the concentration of the peptide with time for low, intermediate and high densities of GO flakes. Normalization was used in order to compare similar initial peptide concentrations.

## Discussions

Our numerical simulations carried out with the commercial finite element software COMSOL Multiphysics (**Fig. 4**) clearly reproduce most of the features of the experimental data (**Fig. 2**). However, after a period of time, the concentration decreases in **Fig. 2**, unlike the monotonically increasing solution to the diffusion equation in **Fig. 4**. The experimental data fits the modeling data if one adds a term to the diffusion equation in order to account for photobleaching, see **Supplementary Fig. S3**. We note that other work on filtration applications of water molecules[5] at the nanometric scale within a GO flake was underpinned by van der Waals interaction[26,27] which played a prominent role at the nanometric scale, but this is not the case in the present work.

## Methods

**Peptide synthesis and diffusion experiments**

Peptides were synthesized by GenScript, based on the previously published peptide sequences[25], with conjugation of FITC fluorophore to the N-terminal of the peptide. A fluorescence plate reader (FLUOstar Optima, BMG Labtech) is used to carry out drug diffusion assays, using excitation filter of 485 nm and an emission filter of 538 nm. The density of graphene oxide is measured by the quantity of GO (ng) deposited onto a total area (mm) of a 24-well culture plate Transwell PET membrane insert and then represented as ng/mm$^3$.

**Transmission electron microscopy**

GO dispersion in water, 4 mg/mL (Sigma-Aldrich) is diluted 100 times with water. 2 µL of the diluted solution is deposited onto an electron microscopy grid. After 20 seconds, the drop is removed with a small fragment of filter paper. A 200 kV with a Tecnai G2 (FEI, Netherlands) and Velata camera (Olympus, Japan) is used for image acquisition.

**Effective model and numerical simulations**

We apply an effective model for porous media to obtain the results in **Fig. 3**. The effective and "free" diffusivities are usually related[28] [according to the equation $D_{eff} = D\varepsilon/\tau$ where ε is the porosity of the structure and τ the tortuosity, which is a measure of the actual length per unit effective length a molecule has to diffuse in a porous structure. To calculate the porosity of the modeled structure we consider the ratio of the perforations to the total computational area, the concentration then obeys Fick's equation[29].

$$\varepsilon \frac{\partial}{\partial t} c - \nabla \cdot (D_{eff} \nabla c) = 0 \qquad (1)$$

Tortuosity is usually expressed as a power of the porosity[30,31] therefore the effective diffusivity varies as $D_{eff} = D\varepsilon^p$.

We note that applying Fick's equation simply leads to an increase in concentration reaching a plateau (steady state) at long times, which does not explain the relative decrease in concentration observed in **Fig. 1d**.

The physical system suffers degradation due to photobleaching. We build in this decay through an additional term $0 < a_{eff} \ll 1$ where $a_{eff}$ is the rate of photobleaching[32-34]

$$\frac{\partial}{\partial t}c - \nabla.(D_{\text{eff}}\nabla c) + a_{\text{eff}}c = 0 \tag{2}$$

We verified that the concentration versus time then follows the same trend as in **Fig. 2**, see **Supplementary Fig. S3**.

AUTHOR INFORMATION

Corresponding Author: Tania M. Puvirajesinghe: tania.guenneau-puvirajesinghe@inserm.fr.

Competing financial interests

The authors declare no competing financial interests.

Author Contributions

T.M.P, Z.L. designed and carried out experiments. R.V.C. and S.G. were responsible for carrying out physical and mathematical approaches and S.G performed COMSOL simulations. All authors have contributed to the writing of this manuscript and given approval to the final version of the manuscript.

ACKNOWLEDGMENTS

The electron microscopy experiments were performed on PiCSL-FBI core facility (IBDM UMR CNRS 7288, Aix-Marseille University), member of the France-BioImaging national research infrastructure. Electron Microscopy analysis was carried out by Fabrice Richard and Jean-Paul Chauvin, with assistance from Aicha Aouane in the Electron Microscopy Platform, IBDML - Institut de Biologie du Développement de Marseille Luminy

UMR 7288, Marseille. Authors would like to thank Samuel Granjeaud for critical comments of the work. The authors wish to acknowledge Vincent Pagneux for useful comments on the diffusion equation to model some degradation effects probably due to photobleaching. This work has been carried out thanks to A*MIDEX project (no ANR-11-IDEX-0001-02) funded by the Investissements d'Avenir French Government program, managed by the French National Research Agency (ANR) with Aix Marseille Université (TMP, A_M-AAP-ID-14-15-140314-09.45-GUENNEAU-PUVIRAJESINGHE-HLS_SAT).

ABBREVIATIONS

GO: graphene oxide; TEM: Transmission Electron Microscopy; PET: polyethylene terephthalate.